\documentclass{article}

\usepackage{microtype}
\usepackage{graphicx}
\usepackage{subcaption}
\usepackage{booktabs} 

\usepackage{hyperref}
\usepackage{esvect}

\usepackage[accepted]{synsml2023}

\usepackage{amsmath}
\usepackage{amssymb}
\usepackage{mathtools}
\usepackage{amsthm}

\usepackage[capitalize,noabbrev]{cleveref}

\theoremstyle{plain}

\theoremstyle{definition}

\theoremstyle{remark}

\synsmltitlerunning{Self-Supervised Deep Learning for Model Correction in the Computational Crystallography Toolbox}

\begin{document}

\twocolumn[
\synsmltitle{Self-Supervised Deep Learning for Model Correction in the Computational Crystallography Toolbox}



\synsmlsetsymbol{equal}{*}

\begin{synsmlauthorlist}
\synsmlauthor{Vidya Ganapati}{yyy,xxx}
\synsmlauthor{Daniel Tcho\'{n}}{yyy}
\synsmlauthor{Aaron S. Brewster}{yyy}
\synsmlauthor{Nicholas K. Sauter}{yyy}
\end{synsmlauthorlist}

\synsmlaffiliation{yyy}{Molecular Biophysics and Integrated Bioimaging Division, Lawrence Berkeley National Laboratory, Berkeley, CA 94720, USA}
\synsmlaffiliation{xxx}{Department of Engineering, Swarthmore College, Swarthmore, PA 19081, USA}

\synsmlcorrespondingauthor{Vidya Ganapati}{vidyag@berkeley.edu}

\synsmlkeywords{Deep Learning, Self-Supervised, Variational Autoencoder, Physics-Informed Deep Learning, Hybrid Modeling, Model Correction}

\vskip 0.3in
]



\printAffiliationsAndNotice{}  

\begin{abstract}

The Computational Crystallography Toolbox (\textsc{cctbx}) is open-source software that allows for processing of crystallographic data, including from serial femtosecond crystallography (SFX), for macromolecular structure determination. We aim to use the modules in \textsc{cctbx} to determine the oxidation state of individual metal atoms in a macromolecule. Changes in oxidation state are reflected in small shifts of the atom's X-ray absorption edge. These energy shifts can be extracted from the diffraction images recorded in serial femtosecond crystallography, given knowledge of a forward physics model. However, as the diffraction changes only slightly due to the absorption edge shift, inaccuracies in the forward physics model make it extremely challenging to observe the oxidation state.  In this work, we describe the potential impact of using self-supervised deep learning to correct the scientific model in \textsc{cctbx} and provide uncertainty quantification. We provide code for forward model simulation and data analysis, built from \textsc{cctbx} modules, at \url{https://github.com/gigantocypris/SPREAD}, which can be integrated with machine learning. We describe open questions in algorithm development to help spur advances through dialog between crystallographers and machine learning researchers. New methods could help elucidate charge transfer processes in many reactions, including key events in photosynthesis. 
\end{abstract}


\section{Introduction}

Crystallography is a branch of science that revolves around investigating the internal structure of crystals. A typical crystal is comprised of a series of almost-identical unit cells, repeated periodically in all directions. Mathematically, it can be described as a convolution of a single average unit cell with a periodic three-dimensional lattice. The distribution of electron density in any crystal can be understood as a three-dimensional periodic wave in direct (experimental) space.

X-ray diffraction (XRD) is a common crystallographic technique used to determine the structure of crystals. In an XRD experiment, incident radiation is scattered by the electron density and is subsequently imaged on a detector. The collected diffraction pattern is related to the Fourier transform of the scatterer density. By the convolution theorem, the Fourier transform is a product of the Fourier transforms of the periodic crystal lattice and the electron density in a single unit cell. In the reciprocal space, this results in a set of discrete peaks that represent the crystal structure. Every peak is indexed using Miller indices $\vv{h}={\left[h \: k \: l \right]}^{\top}$ and carries some information about the crystal structure in the form of a structure factor $F_{\vv{h}}$. Due to the imperfection of real crystals, the peaks in the reciprocal space are not infinitesimally small as expected from theory but rather slightly diffused. This effect can be  modeled by treating the crystal as a finite set of small mosaic domains, each slightly misaligned relative to others. 

For any fixed orientation, a detector images a two-dimensional spherical slice of the reciprocal space called the Ewald sphere. Rotating the crystal in the direct space rotates its Fourier transform in the reciprocal space, allowing the Ewald sphere to pass through different reciprocal space peaks, and deposit the information about their shape and intensity in the form of diffraction spots. The intensity of each diffraction spot, summed incoherently across all mosaic domains, is proportional to the modulus squared of its structure factor $F_{\vv{h}}$. The core problem of XRD structure determination is to retrieve structure factors $F_{\vv{h}}$ based on the intensities of all diffraction spots observed on a detector \cite{giacovazzo_fundamentals_2011}.

The Computational Crystallography Toolbox (\textsc{cctbx}), is open-source software used to process data collected in crystallographic experiments to determine $|F_{\vv{h}}|^2$ of all the diffraction spots in the Fourier transform \cite{grosse-kunstleve_computational_2002}. Documentation and code are available at \url{https://cci.lbl.gov/docs/cctbx} and \url{https://github.com/cctbx/cctbx_project}, respectively. The resulting $|F_{\vv{h}}|^2$ can be processed by another tool such as \textsc{Phenix} \cite{liebschner_macromolecular_2019} in order to solve for the electron density of the macromolecule. The package \textsc{cctbx} models the formation of diffraction images using knowledge of the underlying imaging physics. From the collected data, the inverse problem of finding the underlying structure factor amplitudes can be solved. 

Though the conventional use of \textsc{cctbx} is the determination of structure factor amplitudes, our aim in this work is to use \textsc{cctbx} to determine the oxidation state of individual metal atoms in a macromolecule. Our scientific goal is to understand charge transfers at the atom level in photosystem II, a key protein complex in photosynthesis \cite{bhowmick_structural_2023}. Such knowledge can inform the future development of solar fuels.

Changes in oxidation state can be extracted from the diffraction images recorded in serial femtosecond crystallography (SFX), given knowledge of a forward physics model \cite{sauter_towards_2020}. However, even slight inaccuracies in the forward physics model make it extremely challenging to determine the oxidation state. In this work, we describe the potential impact of using self-supervised deep learning to correct the scientific model in \textsc{cctbx} and provide uncertainty quantification. We provide code for forward model simulation and data analysis, built from \textsc{cctbx} modules, at \url{https://github.com/gigantocypris/SPREAD}, which can be integrated with machine learning. 

We introduce SFX in Section~\ref{sec:sfx} and structure factor amplitude refinement with \textsc{nanoBragg} in Section~\ref{sec:nanobragg}. In Section~\ref{sec:spread}, we describe the use of \textsc{cctbx} and \textsc{nanoBragg} to determine oxidation state of metal atoms with conventional methods. In Section~\ref{sec:model}, we first describe prior work in forward physics model correction and physics-informed variational autoencoders (P-VAEs). We then outline how P-VAEs with model correction can aid in solving for oxidation state. We conclude Section~\ref{sec:model} with open questions in algorithm development to help spur advances through dialog between crystallographers and machine learning researchers. 

\section{Serial Femtosecond Crystallography (SFX)}
\label{sec:sfx}
In a classical diffraction experiment, a single crystal is affixed on a goniometer and cooled down to a cryogenic temperature to limit X-ray radiation damage. Exposure over a series of orientations allows for collection of a complete set of spot intensities. However, the information collected in cryogenic conditions describes a structure far from its natural state. 

\begin{figure}[ht]
     \centering
     \begin{subfigure}[b]{0.23\textwidth}
         \centering
         \includegraphics[width=\textwidth]{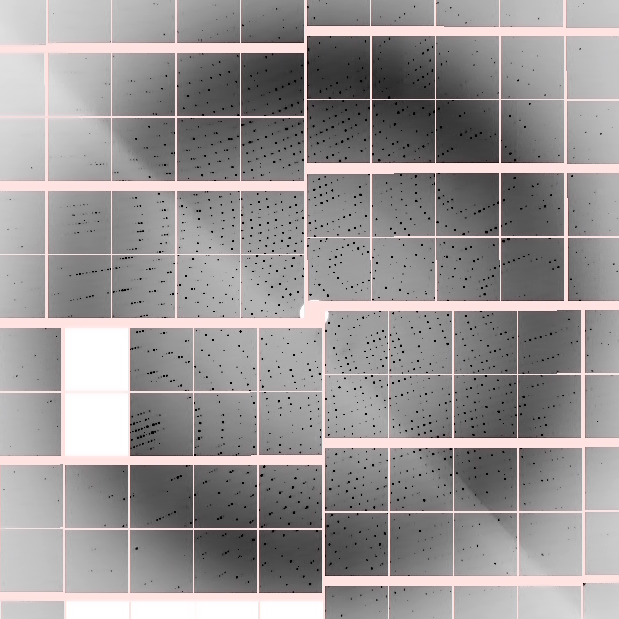}
         \caption{}
         \label{fig:stillshot}
     \end{subfigure}
     \hfill
     \begin{subfigure}[b]{0.23\textwidth}
         \centering
         \includegraphics[width=\textwidth]{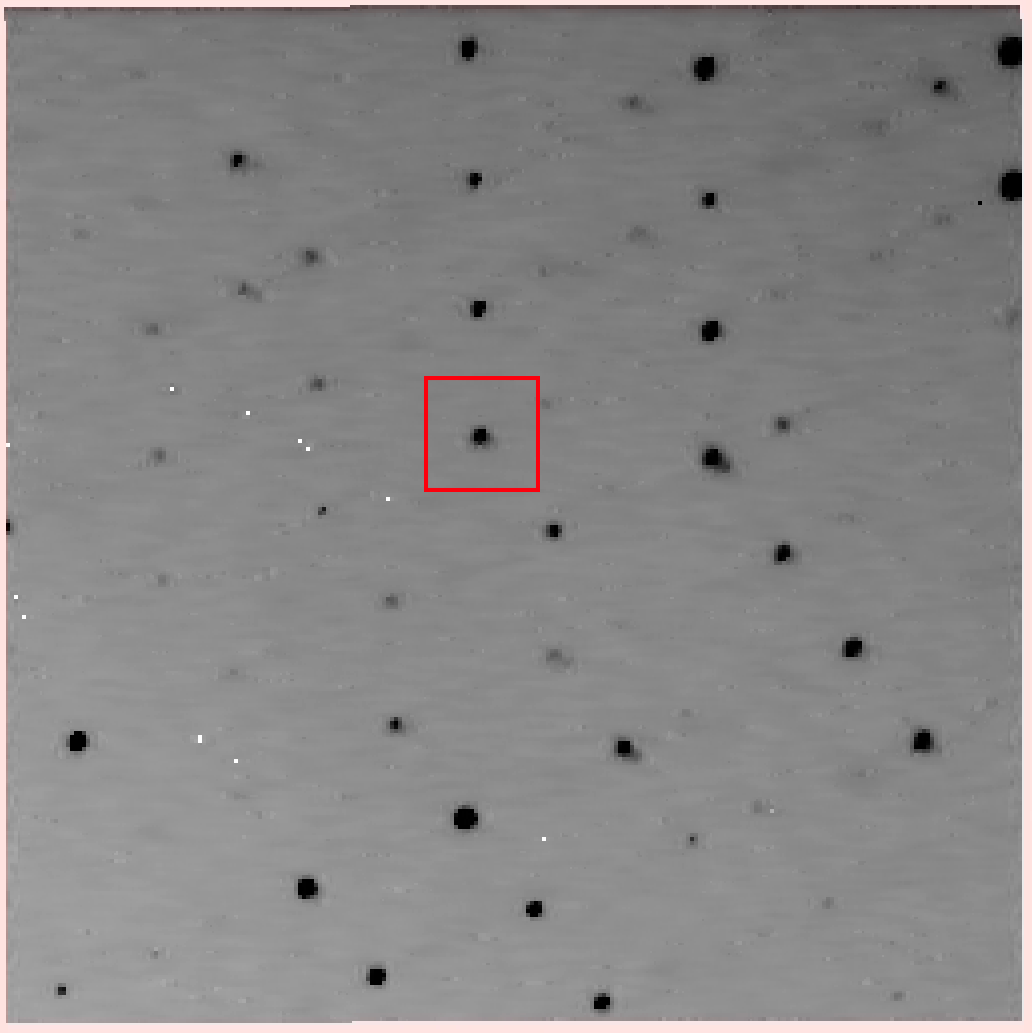}
         \caption{}
         \label{fig:shoebox}
     \end{subfigure}
     \vspace{-1em}
        \caption{(a) Center portion of still shot of photosystem II in an SFX experiment. (b) Example shoebox drawn by \textsc{dials}.}
        \label{fig:exampleSFX}
\end{figure}

Serial femtosecond crystallography (SFX) with X-ray free electron laser (XFEL) pulses allows for room-temperature measurements. In this modality, many micro-crystals are imaged sequentially using short X-ray pulses. A single pulse is used to capture a single ``still shot'', recording a diffraction image of a crystal right before it is destroyed by the radiation. The measurements can be synchronized with an auxiliary laser, allowing observation of short-lived transient states (pump-probe experiments) \cite{kern_room_2012, kern_simultaneous_2013}. All still shots are assumed to originate from crystals with identical chemical composition, but their orientation, exact unit cell lengths, and mosaicity parameters may vary from sample to sample \cite{mendez_beyond_2020}.

Each still shot is composed of pixels collected from multiple detector panels, and the detector is described in a hierarchical manner \cite{brewster_improving_2018}, see Fig.~\ref{fig:stillshot}. The Diffraction Integration for Advanced Light Sources (\textsc{dials}) software package \cite{winter_dials_2018} is built on top of \textsc{cctbx} and can perform diffraction spot finding, see Fig.~\ref{fig:shoebox}. The documentation is available at \url{https://dials.github.io} and the code at \url{https://github.com/dials/dials}. With SFX, the orientation of the crystal in a still shot, as well as an overall scaling factor that depends on variations in the incident beam and volume of illuminated crystal, are initially unknown \cite{evans_how_2013}. The unit cell parameters also vary from crystal to crystal. After spot finding, \textsc{dials} determines the orientation of the crystal and assigns a Miller index to each spot. The structure factor amplitudes are then found by integrating, scaling, and merging the spot intensity values over all still shots \cite{winter_dials_2018}.

\section{Refinement with \textsc{nanoBragg}}
\label{sec:nanobragg}

No crystal is a perfectly periodic monolith; rather, it can be better described as a cluster of small periodic domains. Individual crystalline domains are very similar, but can slightly vary in their orientation and shape. Every diffraction spot originating from the crystal is a sum of the intensity contributions of each mosaic domain. Likewise, the resulting diffraction spots from a polychromatic spectrum can be modeled as the superposition of spots from each wavelength. Both the incident spectrum and mosaicity are not considered in \textsc{dials}, as the software integrates every spot without consideration of its shape. Diffraction spot shape can, however, be modeled with \textsc{nanoBragg}, a module in \textsc{cctbx} \cite{holton_rfactor_2014, lyubimov_advances_2016}. 

During integration, \textsc{dials} investigates a small region of the still shot around each spot called a ``shoebox.'' To solve the inverse problem of determining the structure factors as well as other parameters such as orientation and unit cell, the shoeboxes can be simulated from the underlying parameters with \textsc{nanoBragg}. The simulation can be then compared with the experimental data to determine a loss function. The structure factors and shoebox-dependent parameters can be updated by the gradients with respect to a loss function until convergence \cite{mendez_beyond_2020}. The structure factor and parameter estimates from \textsc{dials} are used as initial conditions to make the inverse problem computationally tractable. We note that with both \textsc{dials} and \textsc{nanoBragg}, structure factor amplitudes are point estimates; there is no uncertainty quantification.

\section{Spatially Resolved Anomalous Dispersion}
\label{sec:spread}

Building off of \textsc{nanoBragg} and \textsc{cctbx}, we can determine the oxidation state of individual metal atoms in a macromolecule with data from SFX. Changes in oxidation state are reflected in slight shifts (on the order of 1-2 electron volts) of the atom's \textit{K} absorption edge. These shifts are embedded in SFX data. There is a wavelength dependency in the structure factor $F_{\vv{h}}$ as the scattering factor of each atom in the macromolecule includes a complex wavelength-dependent quantity known as the anomalous scattering factor \cite{sauter_towards_2020}. Far from the \textit{K} absorption edge, the scattering factor is approximately constant over wavelength. We aim to solve for the anomalous scattering factor for atoms with $K$ edge near the center wavelength of the incident spectrum. This technique, known as spatially resolved anomalous dispersion (SPREAD), can yield insight into electron movement during a chemical reaction. In particular, our scientific aim is to solve for the anomalous scattering factors of the four manganese (Mn) atoms in photosystem II to elucidate single-electron transfers in photosynthesis \cite{sauter_towards_2020}. 

The structure factor at the Miller index $\vv{h}$ is given as the sum of contributions from each atom $m$ of the macromolecule:

\begin{equation}
F_{\vv{h}}(\lambda) = \sum_{m} F_{\vv{h},m}(\lambda),
\end{equation}

where $\lambda$ denotes the wavelength of incident radiation. As described in \cite{sauter_towards_2020}, the wavelength dependent contribution from each atom, $F_{\vv{h},m}$, can be expressed as:

\begin{equation}
\begin{aligned}
F_{\vv{h},m}(\lambda) = {} & \left[ f_m^0 (|\vv{Q}|) + \Delta f^{\prime}_m (\lambda) + i \Delta f^{\prime \prime}_{m} (\lambda ) \right] \times \\
		& \exp \left[ 2 \pi i \vv{r_m} \cdot \vv{h} \right] \times \exp (-B_m |\vv{Q}|^2 / 4),
\end{aligned}
\end{equation}

where $\Delta f^{\prime}_m$ and $\Delta f^{\prime \prime}_{m}$ are the real and imaginary parts of the wavelength-dependent anomalous scattering factor for the ${m}^\textrm{th}$ atom, related by the Kramer's Kronig relationship \cite{meurer_refinement_2022, sherrell_diffraction_2014}. The anomalous scattering factor changes with valence state, but is constant over the magnitude of the scattering vector $|\vv{Q}|$. The non-anomalous scattering factor of the atom is denoted by $f_m^0$ and depends on the scattering vector $|\vv{Q}|$. The position of the atom within the unit cell using fractional coordinates is $\vv{r_m}$. The Miller index is denoted by $\vv{h}$, while $B_m$ is the atom's temperature-dependent $B$ factor. For an incident spectrum centered at 6550 eV, the $\Delta f^{\prime}_m$ and $\Delta f^{\prime \prime}_{m}$ terms are negligible for all photosystem II atoms, except the four Mn atoms. The functions $\Delta f^{\prime}_m (\lambda)$ and $\Delta f^{\prime \prime}_{m} (\lambda)$ shift by a few electron volts between manganese in its $3+$ and $4+$ oxidation state; this is the change we aim to determine.

The change in the total structure factor due to a change in the electronic state of a few constituent atoms is small. There are thus strict requirements on the accuracy of the forward physics model. For example, an inaccurate description of crystal mosaicity may lead to an incorrect determination of the anomalous scattering factors. So far, SPREAD with SFX data has only been performed successfully with simulated data, with code that extends \textsc{nanoBragg} and \textsc{cctbx} \cite{sauter_towards_2020}. This prior work on simulated data results in point estimates of the anomalous scattering factor as a function of wavelength for the atoms of interest. Application of the methods to real SFX data has not been successful thus far. We describe the potential impact of using self-supervised deep learning to correct the scientific model and provide uncertainty quantification. 

\begin{figure*}[ht]
\centering
\includegraphics[width=0.8\textwidth]{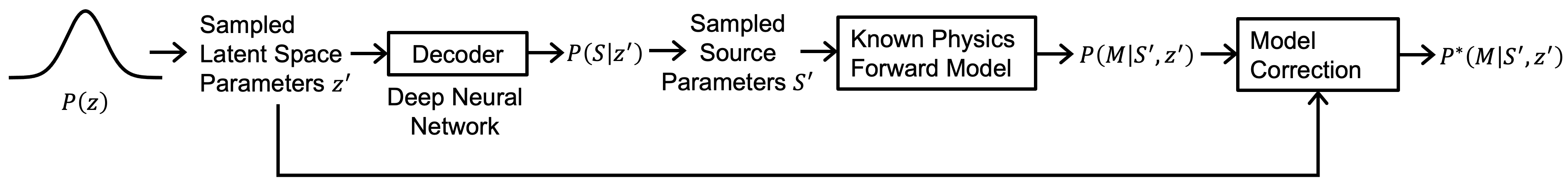}
\caption{Generator for source parameters; model correction could be performed with a normalizing flow \cite{kobyzev_normalizing_2021}.}
\label{fig:generator}
\end{figure*}

\section{Model Correction with Neural Networks}
\label{sec:model}

In an inverse problem, we have measurements and a known forward physics model. Our aim is to discover the source of those measurements. If we know the source and the forward physics model (the forward problem), determining the probability distribution of measurements is straightforward. However, the inverse problem of determining the source given the measurements is more challenging and may be ill-posed. If we have a fully specified forward model, we can take an optimization-based approach, using gradient descent to tune an initial guess of the source in order to maximize the likelihood of achieving the given measurements. The inverse problem becomes even more challenging if we have incomplete or incorrect knowledge of the forward model. The complete, correct form of the forward model may be unknown due to experimental unknowns and complexities. If we have measurements on multiple sources, with each measurement obeying the same underlying (incomplete) forward model, we may have enough information to both correct the model and determine all the sources. In this case, our optimization objective is to maximize the total likelihood of all measurements, with a penalty for deviating too far from the known forward physics. Here, we describe related work, outline our framework for model correction in SFX, and discuss open questions.


\begin{figure*}[ht]
\centering
\includegraphics[width=0.9\textwidth]{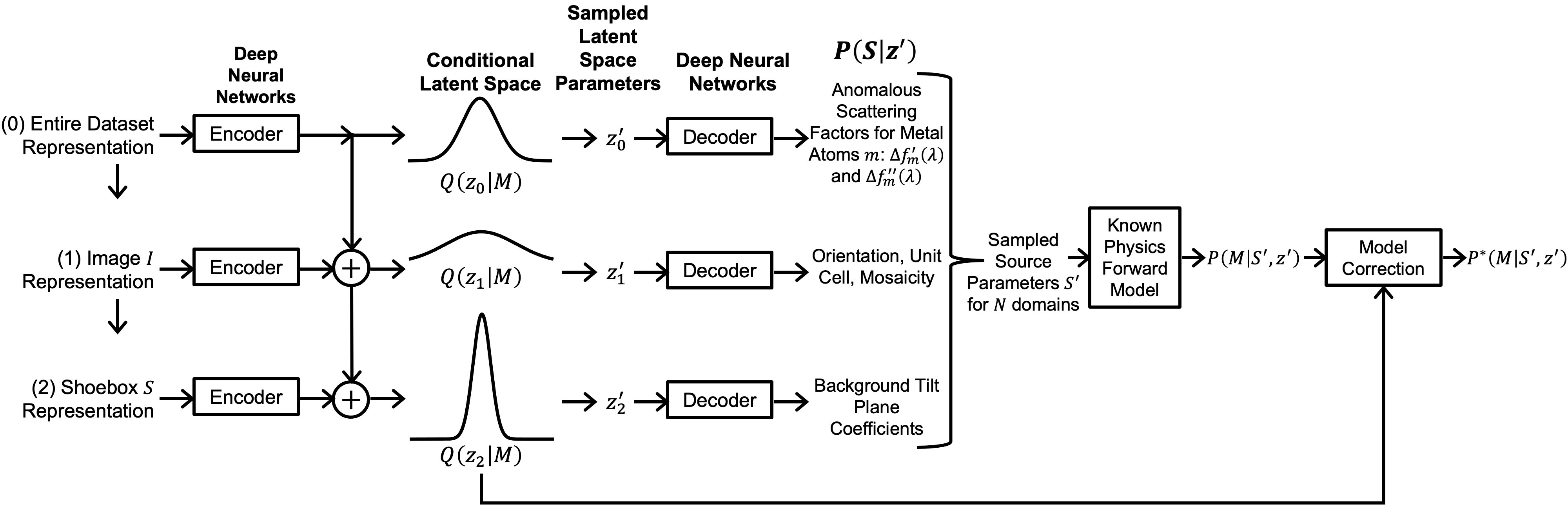}
\caption{Hierarchical P-VAE for determining anomalous scattering factors.}
\label{fig:pvae}
\end{figure*}

\subsection{Related Work}

\subsubsection{\textsc{Careless}}

The software package \textsc{Careless} \cite{dalton_unifying_2022} provides a model-free way to correct structure factor amplitudes $|F_{\vv{h}}|$ derived from \textsc{dials}. Each structure factor from \textsc{dials} is assumed to be the product of an image-dependent, spot-dependent scale factor, and the true structure factor amplitude. The prior distribution on the true structure factor amplitude is given; the prior on the scale factor is assumed to be uninformative. A neural network takes metadata on the diffraction spot (e.g. crystal orientation, location on the detector, image number, and Miller index), and outputs a prediction of the posterior probability distribution of the corresponding scale factor. Variational inference is used to train this  neural network as well as the parameters of the distribution estimating the structure factor amplitude posterior, pushing both distributions to the true posteriors. Intuitively, variational inference attempts a balance between maximizing the likelihood of the measured intensity data while not veering far from the prior distribution on the structure factor $|F_{\vv{h}}|$. The metadata used as input to the neural network must be chosen judiciously, as it is possible for the scale factor to overexplain the experimental $|F_{\vv{h}}|$, creating a poor true structure factor amplitude estimate. 

Like \textsc{Careless}, we aim to perform model correction. However, our goal is primarily to find anomalous scattering factors of certain atoms; for SPREAD, we assume the structure factor amplitude is known. For this task, our framework needs to take into account pixel-by-pixel variations in diffraction spot shoeboxes, as well as utilize the (partially) known underlying physics.

\subsubsection{Physics-Informed Variational Autoencoder (P-VAE)}

Recent work has evaluated the use of physics-informed variational autoencoders (P-VAEs) to solve inverse problems in imaging \cite{mendoza_self-supervised_2022, olsen_data-driven_2022}; a similar formulation is described in \cite{leong_ill-posed_2023}. Specifically, these works consider a dataset of measurements, with each measurement on a different unknown source. Due to experimental limitations, the measurement on each source is sparse. There is not enough information in a single measurement $M$ to recover the corresponding source $S$ using conventional optimization methods for inverse problems. However, the entire dataset of measurements is large, i.e. there are sparse measurements on many similar sources. The P-VAE jointly solves for the underlying prior distribution on the sources $P(S)$ and all the posterior distributions on the dataset, $P(S | M)$. In SFX, we have a similar problem where we have sparse (i.e. single orientation) measurements on many similar but different crystals. The formulation of the P-VAE can assist in the determination of a probability distribution for the anomalous scattering factors, as opposed to yielding just a point estimate.

\subsubsection{Incomplete Forward Models}

The frameworks in \cite{mendoza_self-supervised_2022, olsen_data-driven_2022, leong_ill-posed_2023} focus on the problem of sparse measurements, they do not consider an incomplete forward model. A partially specified forward physics model in the context of a P-VAE is considered in \cite{takeishi_physics-integrated_2021}. A neural network is trained to transform the incomplete model into the completed one. The augmented forward model is penalized for veering from the known incomplete forward model through additional loss terms added to the P-VAE loss. To solve for the anomalous scattering factors in SFX, we face the dual problems of sparse measurements and an incomplete forward model.

\subsection{P-VAE for SPREAD}

Here, we outline how P-VAEs could be applied to processing SFX data for SPREAD and describe open questions. Code and documentation for the SPREAD forward model are given at \url{https://github.com/gigantocypris/SPREAD}. 

We process collected still shots with \textsc{dials}, drawing shoeboxes around diffraction spots and obtaining estimates for unit cell shape, orientation, and overall scale factor. We aim to find a probability distribution of the underlying anomalous scattering factor functions $\Delta f^{\prime}_m (\lambda)$ and $\Delta f^{\prime \prime}_{m} (\lambda)$. To do so, we can create a ``generator'' with a latent space $z$ that can be sampled to yield shoebox source parameters from the same underlying distribution as the measured shoeboxes. The generator includes the partially known forward physics model and model correction; see Fig.~\ref{fig:generator}. 

We want to train the generator to maximize the probability of obtaining the actual shoeboxes. This problem is made computationally tractable by using an encoder that takes an actual shoebox and its metadata as input, and approximates the conditional probability distribution $P(z | \text{shoebox})$. Connecting the encoder to the generator creates a P-VAE. The derivation of the P-VAE loss function with model correction is given in Appendix~\ref{appendix_derivation}. Training the networks with the P-VAE loss recovers the distribution governing the anomalous scattering factors $\Delta f^{\prime}_m (\lambda)$ and $\Delta f^{\prime \prime}_{m} (\lambda)$.

The data of a single shoebox can be represented in a hierarchical manner, with latent parameters shared amongst all shoeboxes at the dataset and image levels. We outline the basic framework of a hierarchical P-VAE in Fig.~\ref{fig:pvae}, with further details in Appendix~\ref{appendix_derivation}. We describe how a similar framework can be used to refine structure factor amplitudes in Appendix~\ref{appendix_structure}.

\subsection{Open Questions}

We describe a potential framework to use machine learning for model correction in SFX. However, there are many open questions, such as:

\begin{itemize}
  \item What trade-off do we make between following the scientific model generated by first principles and allowing model corrections with deep neural networks? Relatedly, how do we verify correctness?
  \item Conventional crystallographic data analysis rejects a significant portion of collected data. Can deep learning techniques allow for insight to be extracted from poor-quality data?
  \item What neural network architecture is needed for model correction? How sensitive is the procedure to neural network architecture?
  \item How do we best incorporate the results from \textsc{dials} and \textsc{nanoBragg} for quantities such as unit cell, orientation, scale, and mosaicity?
\end{itemize}

\section{Conclusion}

We describe the Computational Crystallography Toolbox (\textsc{cctbx}) and the potential for applying self-supervised physics-informed deep learning methods for analysis in serial femtosecond crystallography (SFX). A scientific problem of interest in SFX is determining the anomalous scattering factors of specific metal atoms in macromolecules. Such knowledge will allow determination of the oxidation state of individual metal atoms in the macromolecule, elucidating charge transfer processes in chemical reactions. 

We outline the potential use of deep neural networks to make arbitrary corrections to the forward model in SFX, supplementing previous work \cite{sauter_towards_2020, mendez_beyond_2020, brehm_crystal_2023} that solely optimizes variables that parametrize a forward model derived from first principles. The goal is to correct for experimental effects of unknown origin, relaxing the stringent model accuracy requirements for spatially resolved anomalous dispersion (SPREAD). These methods have the potential to improve data analysis, with the impact of discovering new science by striking a balance between knowledge of an ideal forward model, and knowledge learned from data. We present open questions to facilitate collaborations between crystallographers and deep learning researchers, with the aim of accelerating progress in SFX. Code for forward model simulation with \textsc{cctbx} modules and instructions for conventional analysis with \textsc{dials} are given at \url{https://github.com/gigantocypris/SPREAD}.

\section*{Acknowledgements}

The authors acknowledge Daniel W. Paley and Iris D. Young for useful discussions and the still shot of photosystem II. V.G. acknowledges support from the National Science Foundation CAREER (grant 2236796). N.K.S. acknowledges support from the National Institutes of Health (grant R01-GM117126), and the Exascale Computing Project (grant 17-SC20-SC), a collaborative effort of the DOE Office of Science and the National Nuclear Security Administration. 

\bibliography{paper}
\bibliographystyle{synsml2023}

\newpage
\appendix
\onecolumn

\section{Derivation of Physics-Informed Variational Autoencoder (P-VAE) Loss}
\label{appendix_derivation}

\begin{figure}[ht]
\centering
\includegraphics[width=0.4\textwidth]{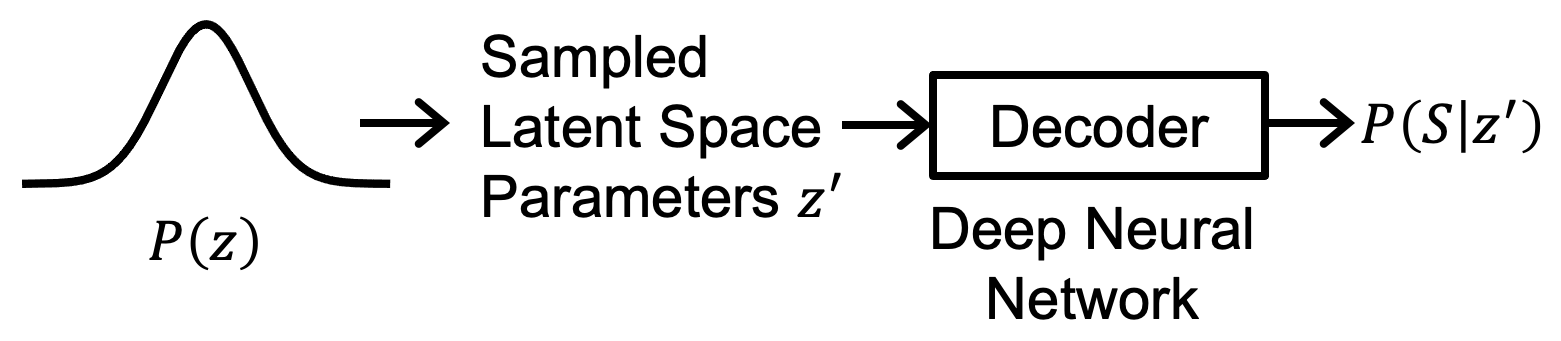}
\caption{Generator for sources $S$.}
\label{fig:bgenerator}
\end{figure}

In a variational autoencoder \cite{kingma_auto-encoding_2014, doersch_tutorial_2016}, the goal is to learn how to generate new examples, sampled from the same underlying probability distribution as a training dataset of $m$ sources $\{S_1, S_2, ... S_m\}$. In our case, a single source $S$ fully specifies a diffraction spot shoebox with underlying parameters including orientation, unit cell, mosaicity, anomalous scattering factors of metal atoms, parameters characterizing the background, and model correction terms. To accomplish the task of creating a shoebox generator, a latent random variable $z$ is created that describes the space on a lower-dimensional manifold. A deep neural network defines a function (the ``decoder'') from a sample of $z$ to a conditional probability distribution $P(S | z)$, see Fig.~\ref{fig:bgenerator}. 

The parameters of the decoder network are optimized to maximize the probability of generating the independent sources of the training dataset: 

\begin{equation}
\log P(S_1, S_2, ... S_m) = \sum_{i=1}^m \log P(S_i) = \sum_{i=1}^m \log \left[  \int P(S_i | z) P(z) dz \right]
\end{equation}

However, we do not have any ground truth sources $S$, but rather a dataset of noisy diffraction spot shoebox measurements $\{M_1, M_2, ... M_m\}$. Each measurement $M$ consists of the pixel intensity values inside the shoebox. If we assume the forward model $P(M | S)$ is known, instead of maximizing the probability of generating $S$, we can maximize the probability of generating the measurements: 

\begin{equation}
\log P(M_1, M_2, ... M_m) = \sum_{i=1}^m \log P(M_i) = \sum_{i=1}^m \log \left[  \int \int P(M_i | S) P(S | z) P(z) dS dz \right] 
\end{equation}

\begin{figure}[ht]
\centering
\includegraphics[width=0.7\textwidth]{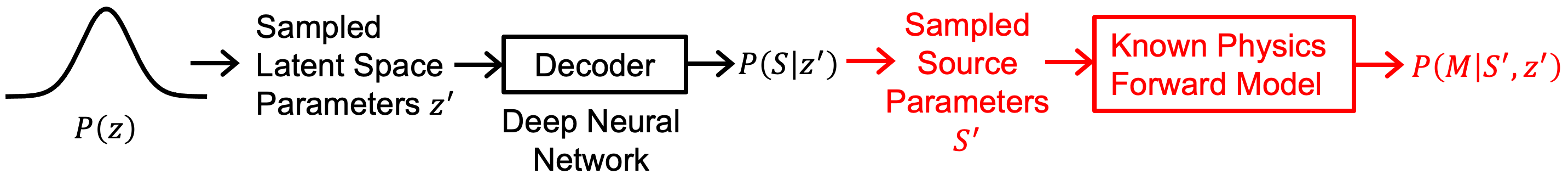}
\caption{Physics-informed generator for sources $S$. The modifications to a conventional generator are highlighted in red.}
\label{fig:pgenerator}
\end{figure}

This modified ``physics-informed'' generator is seen in Fig.~\ref{fig:pgenerator}. 

If the forward model is only partially specified, the likelihood $P(M | S)$ can be modified to $P^{*}(M | S)$ by processing through a normalizing flow \cite{kobyzev_normalizing_2021}, see Fig.~\ref{fig:generator}. 


A penalty for deviating from the known forward model can be added to the overall optimization objective, such as a term proportional to the Kullback–Leibler (KL) divergence between the modified and unmodified likelihoods, to force the generator to try to first explain the data with physics before applying a trainable modification. 

\begin{figure}[ht]
\centering
\includegraphics[width=0.9\textwidth]{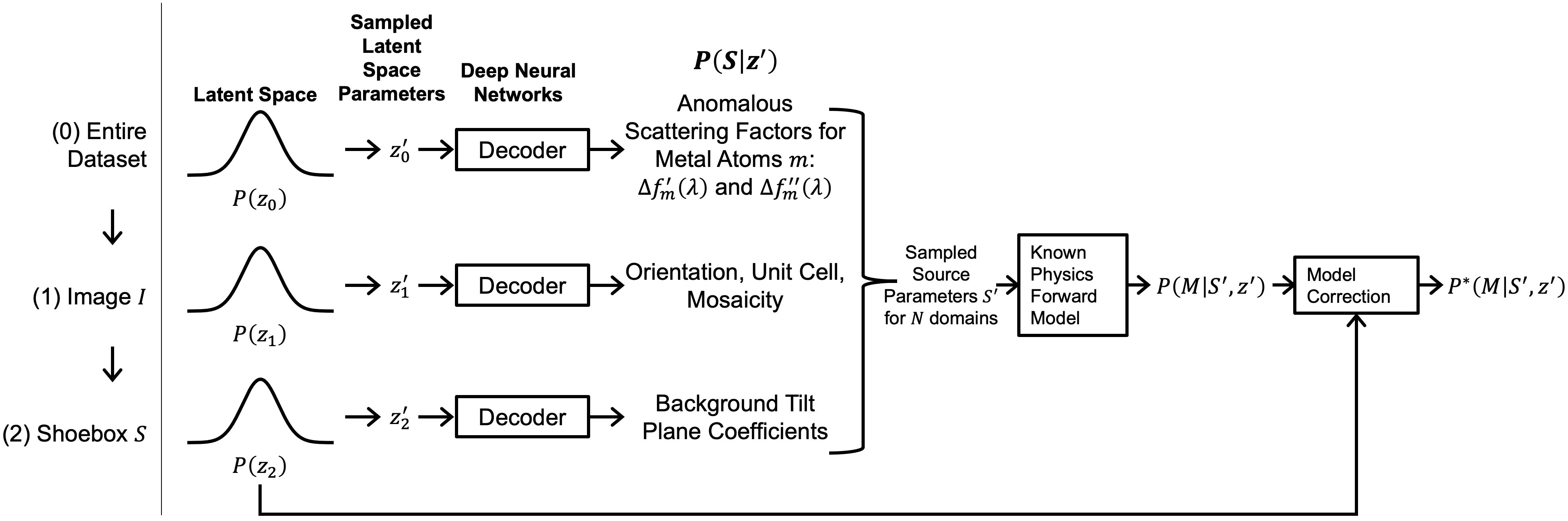}
\caption{Hierarchical physics-informed generator with model correction for sources $S$.}
\label{fig:hgenerator}
\end{figure}

We note that different independent latent variables can separately underlie quantities such as unit cell, mosaic shape and size, and orientation. The latent variables can be organized hierarchically: global over the entire dataset (e.g. anomalous scattering factors), per image (e.g. orientation, incident photon spectrum), and per shoebox (e.g. background noise parameters). The crystal is composed of many mosaic domains, each with a different unit cell, shape, size, and mis-orientation; the set of source parameters can be sampled $N$ times, where $N$ is the number of total mosaic domains modeled. The hierarchical generator can output shoeboxes with knowledge of the pixel positions on the detector, see Fig.~\ref{fig:hgenerator}.

\begin{figure}[ht]
\centering
\includegraphics[width=0.4\textwidth]{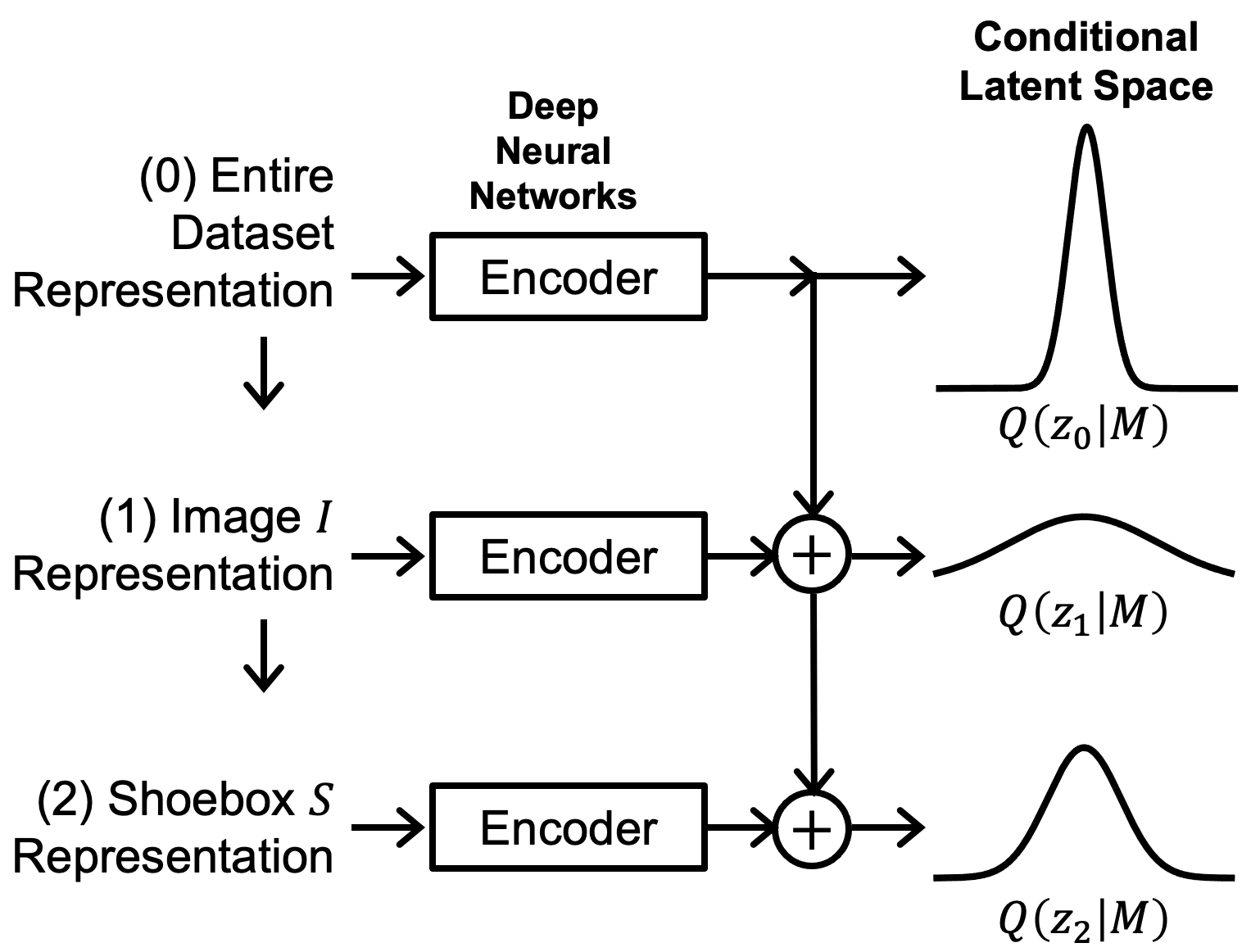}
\caption{Hierarchical encoder for a P-VAE.}
\label{fig:encoder}
\end{figure}

Considering a single example (as in stochastic training with batch size of $1$), we aim to maximize $\log P(M)$; for most sampled values of $z^{\prime}$ and $S^{\prime}$, the probability $P(M | S^{\prime}, z^{\prime})$ is close to zero, causing poor scaling of sampled estimates to the integral $\int \int P(M | S) P(S | z) P(z) dS dz$. We follow the P-VAE formulation \cite{mendoza_self-supervised_2022, olsen_data-driven_2022}, estimating the parameters of $P(z|M)$ by processing the measurement $M$ using a deep neural network called the ``encoder,'' see Fig.~\ref{fig:encoder}.

The output of the encoder is an estimate of $P(z|M)$ denoted as $Q(z|M)$. The KL divergence between the distributions is given by:
\begin{equation}
D_{KL}[Q(z|M) || P(z|M)] = E_{z \sim Q}[\log Q(z|M) - \log P(z|M)]
\end{equation}

We also have, by Bayes' Theorem: 
\begin{equation} 
\log P(z|M) = \log P(M|z) + \log P(z) - \log P(M)
\end{equation}
 
Combining the expressions yields: 
\begin{equation}
\log P(M) - D_{KL}[Q(z|M) || P(z|M)] = E_{z \sim Q} \left[ \log \int P(M | S)P(S|z)dS\right] - D \left[ Q(z|M) || P(z) \right].
\end{equation}

The first term on the right side of this expression can be estimated with a sample-based estimate. As KL divergence is always $\geq 0$ and reaches $0$ when $Q(z|M) = P(z|M)$, maximizing the right side during training causes $P(M)$ to be maximized while forcing $Q(z|M)$ towards $P(z|M)$. When forward model correction is applied, a term can be added to penalize the distance between $P(M | S)$ and $P^{*}(M | S)$. The full physics-informed variational autoencoder is visualized in Fig.~\ref{fig:pvae}.

\section{Framework for Structure Factor Refinement}

\label{appendix_structure}

\begin{figure}[ht]
\centering
\includegraphics[width=0.9\textwidth]{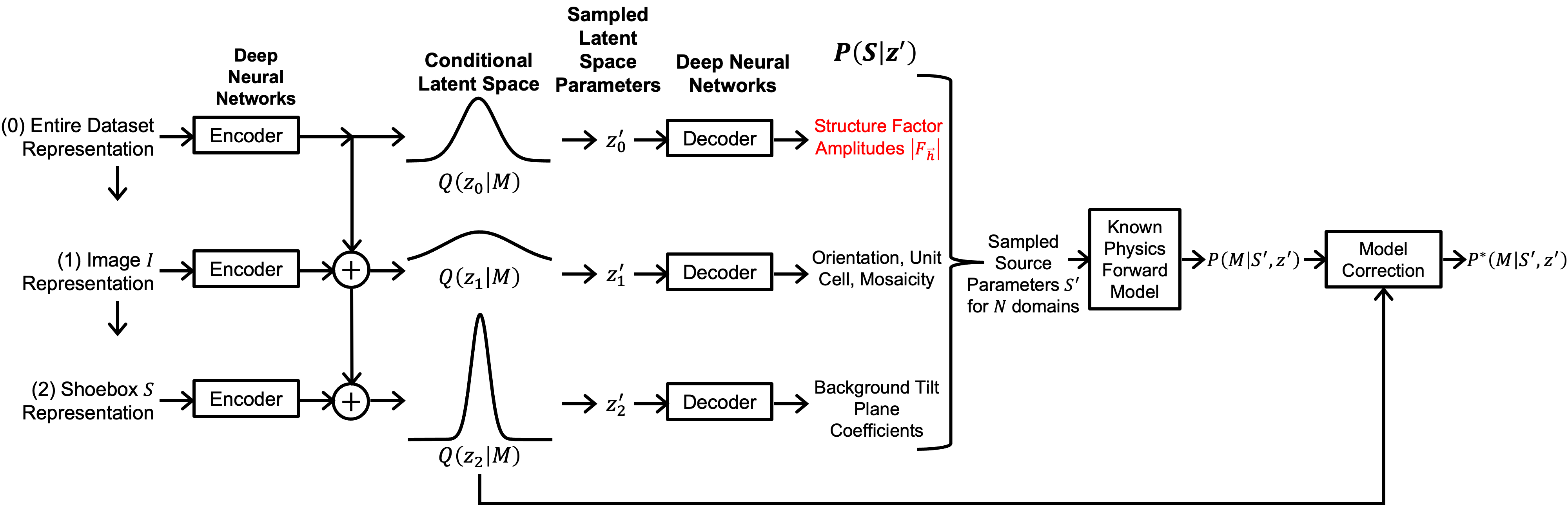}
\caption{Hierarchical P-VAE for determining structure factor amplitudes.}
\label{fig:pvae_s}
\end{figure}

Our focus in this paper is the determination of anomalous scattering factors. Machine learning has great potential for impact in this area, as there are stringent forward model accuracy requirements. In the solution of anomalous scattering factors, we assume knowledge of the structure of the macromolecule. However, refinement of the structure factor amplitudes is also possible from a similar framework, see Fig.~\ref{fig:pvae_s}. 

Initial estimates of structure factor amplitude as well as orientation and unit cell can be found with \textsc{dials} and \textsc{nanoBragg}. These values can be used for initial training of the encoder and decoder, allowing a warm-start to the training procedure.

\end{document}